\documentclass{PoS}

\title{Invariant yield and azimuthal anisotropy measurements of strange and multi-strange hadrons in Au+Au collision at $\sqrt{s_{NN}}$ = 27 and 54.4 GeV at STAR}

\ShortTitle{spectra and flow}

\author{\speaker{Prabhupada Dixit (for the STAR collaboration)}\thanks{A footnote may follow.}\\
        Indian Institute of Science Education and Research, Berhampur, India\\
        E-mail: \email{dixitprabhupada471@gmail.com}}

\abstract{ In this proceedings, we have presented the invariant yield measurements of strange and multi strange hadrons such as $K^{0}_{S}$, $\Lambda$, $\bar{\Lambda}$, $\phi$, $\Xi^{-}$, $\bar{\Xi}^{+}$, $\Omega^{-}$, $\bar{\Omega}^{+}$ in Au+Au collisions at $\sqrt{s_{NN}}$ = 54.4 GeV at midrapidity ($|y|<0.5$). The second and third-order azimuthal anisotropic flow of the above mentioned particles at $\sqrt{s_{NN}}$ = 27 and 54.4 GeV are also discussed. Some of the important observables such as nuclear modification factor ($R_{CP}$), baryon to meson ratio ($\Omega/\phi$), transverse momentum dependence of $v_{2}$ and $v_{3}$  are measured. The number of constituent quark (NCQ) scaling of $v_{2}$ and $v_{3}$ is studied for all these particles.

}

\FullConference{International conference on Critical Point and Onset of Deconfinement (CPOD)\\
                15-19 March 2021\\
                Online}

\begin{document}

\section{Introduction}
Quantum chromodynamics predicts the existence of a deconfined state of quarks and gluons at high temperature and density. Heavy-ion collisions are a unique tool to create and study such a state of the matter. There are many observables to probe the QGP phase of the system created in heavy-ion collisions, some of them are nuclear modification factor, baryon to meson ratio, anisotropic flow.
\par
The nuclear modification factor is the ratio of the particle yield in central to peripheral collisions scaled by the number of binary collisions which is given by
\begin{equation}
\label{eq-2}
R_{CP} = \frac{[\frac{dN}{dp_{T}}\frac{1}{N_{coll}}]_{central}}{[\frac{dN}{dp_{T}}\frac{1}{N_{coll}}]_{peripheral}}.
\end{equation}
Suppression of the ratio at high $p_{T}$ indicates the energy loss by the high $p_{T}$ partons inside colour field of the QGP medium. The baryon to meson ratio, $\Omega/\phi$, gives us information about the hadron formation mechanism during the hadronization. Another important observable to probe the QGP phase is anisotropic flow. Flow is defined as the collective expansion of the medium created in heavy-ion collisions. Depending on the initial spatial geometry and event-by-event fluctuations in the distribution of nucleons inside the colliding nuclei there can be the presence of different orders of anisotropy in the flow. It has been predicted by model calculations that anisotropic flow coefficients are sensitive to the equation of the state and shear viscosity to entropy density ratio ($\eta/s$)~\cite{Ref1} . To study the flow, we use Fourier series expansion of the azimuthal distribution of the invariant yield given by
\begin{equation}
\label{eq-1}
E\frac{d^{3}N}{dp^{3}} =\frac{1}{2\pi}\frac{d^{2}N}{p_{T}dp_{T}dy}\left[1 + \sum_{n} 2v_{n}\cos n(\phi-\Psi_{R})\right].
\end{equation}
Here, $v_{1}$, $v_{2}$ and $v_{3}$ represent directed flow, elliptic flow and triangular flow respectively. The $n^{th}$-order flow coefficient is given by 
\begin{equation}
\label{eq-2}
v_{n} = \langle \cos n(\phi -\psi_{n})\rangle,
\end{equation}
where, $\psi_{n}$ is the $n^{th}$ order event plane. A detailed description of event plane and its calculation can be found in Ref.~\cite{Refa}.
\par
We have used strange and multi-strange hadrons in our study since they are expected to freeze out earlier and have small hadronic interaction cross section~\cite{Ref2} compared to other charged hadrons. So they get least affected by the late-stage hadronic interaction phase of the system.
\section{Analysis details}
Since the particles used in our analysis are short-lived and decay before reaching the detector, we have reconstructed them by using the invariant mass of their daughter particles. The decay channels and branching ratios of these particles are given below.\\

\begin{figure}[h]
\centering
\includegraphics[width=7cm,clip]{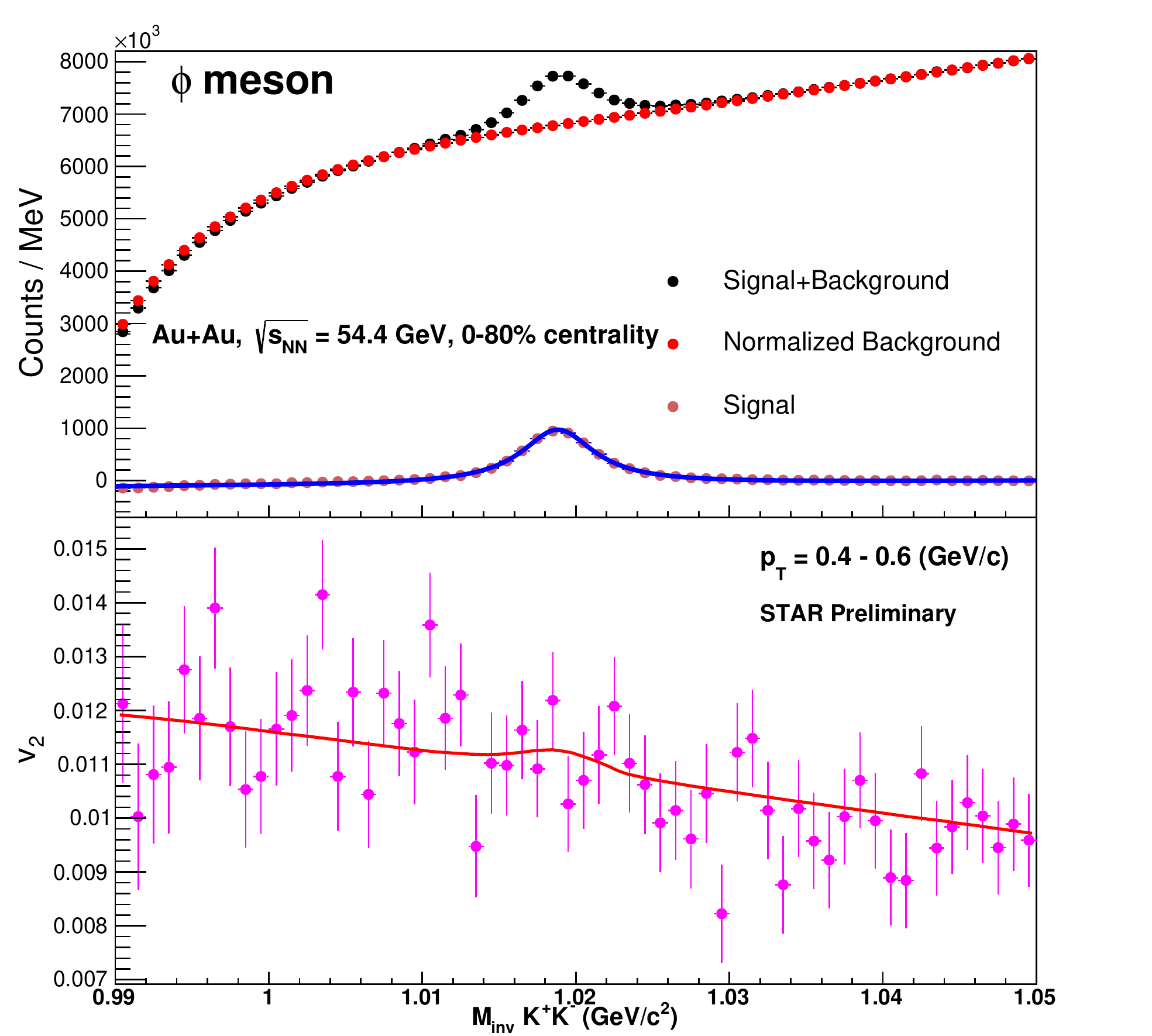}
\caption{The upper panel shows the signal+background distribution and signal peak of phi meson in Au+Au collision at $\sqrt{s_{NN}}$ = 54.4 GeV in $p_{T}$ bin 0.4-0.6 GeV/$c$. The lower panel shows $v_{2}$ as a function of invariant mass of $K^{+}$ and $K^{-}$. The distribution is fitted with a function given in Eq.~\ref{Eq-2}. }
\label{fig-2}       
\end{figure}

\begin{center}
$K_{S}^{0} \rightarrow \pi^{+} + \pi^{-}$ (B.R = 69.20\%)\\
$\Lambda(\bar{\Lambda}) \rightarrow p(\bar{p}) + \pi^{-}(\pi^{+})$ (B.R = 63.9\%)\\
$\Xi(\bar{\Xi^{+}}) \rightarrow \Lambda(\bar{\Lambda}) + \pi^{-}(\pi^{+})$ (B.R = 99.887\%)\\
$\Omega(\bar{\Omega^{+}}) \rightarrow \Lambda(\bar{\Lambda}) + K^{-}(K^{+})$ (B.R = 67.8\%)\\
$\phi \rightarrow K^{+} + K^{-}$ (B.R = 49.1\%)
\end{center}
For particles like $K_{S}^{0}$, $\Lambda$, $\Xi$ and $\Omega$ weak decay topology cuts are applied to reduce the combinatorial background. For $\Lambda$ hyperons weak decay feed-down correction is implemented. In the upper panel of Fig.~\ref{fig-2}, the invariant mass distribution is shown for $K^{+}$ and $K^{-}$ to reconstruct the $\phi$ mesons as an example. The combinatorial background is calculated by event-mixing method and subtracted from the signal+background distribution to get the signal of $\phi$ mesons.



For flow coefficient measurements we have used invariant mass method~\cite{Ref3}. 
As an example we have explain $v_{2}$ calculation of $\phi$ mesons using this method. In this method we need to calculate $v_{2}$ as a function of invariant mass of decay daughters, $K^{+}$ and $K^{-}$ as shown in the lower panel of Fig.~\ref{fig-2}. The distribution is fitted with a function given in Eq.~\ref{Eq-2}, where $v_{2}^{S+B}$ is the combined $v_{2}$ of signal+background. $v_{2}^{S}$ and $v_{2}^{B}$ represent $v_{2}$ of signal and background respectively. $v_{2}^{B}$ is parametrized as a first order polynomial of invariant mass. $v_{2}^{S}$ is a free parameter and can be extracted from the fitting. Same method was used to calculate $v_{2}$ and $v_{3}$ of other particles as well.
\begin{equation}
v_{2}^{S+B} (M_{inv})= v_{2}^{S} \frac{S}{S+B}(M_{inv}) + v_{2}^{B}\frac{B}{S+B}(M_{inv}).
\label{Eq-2}
\end{equation}

\section{Results and Discussion}

\begin{figure}[h]
\centering
\includegraphics[width=9cm,clip]{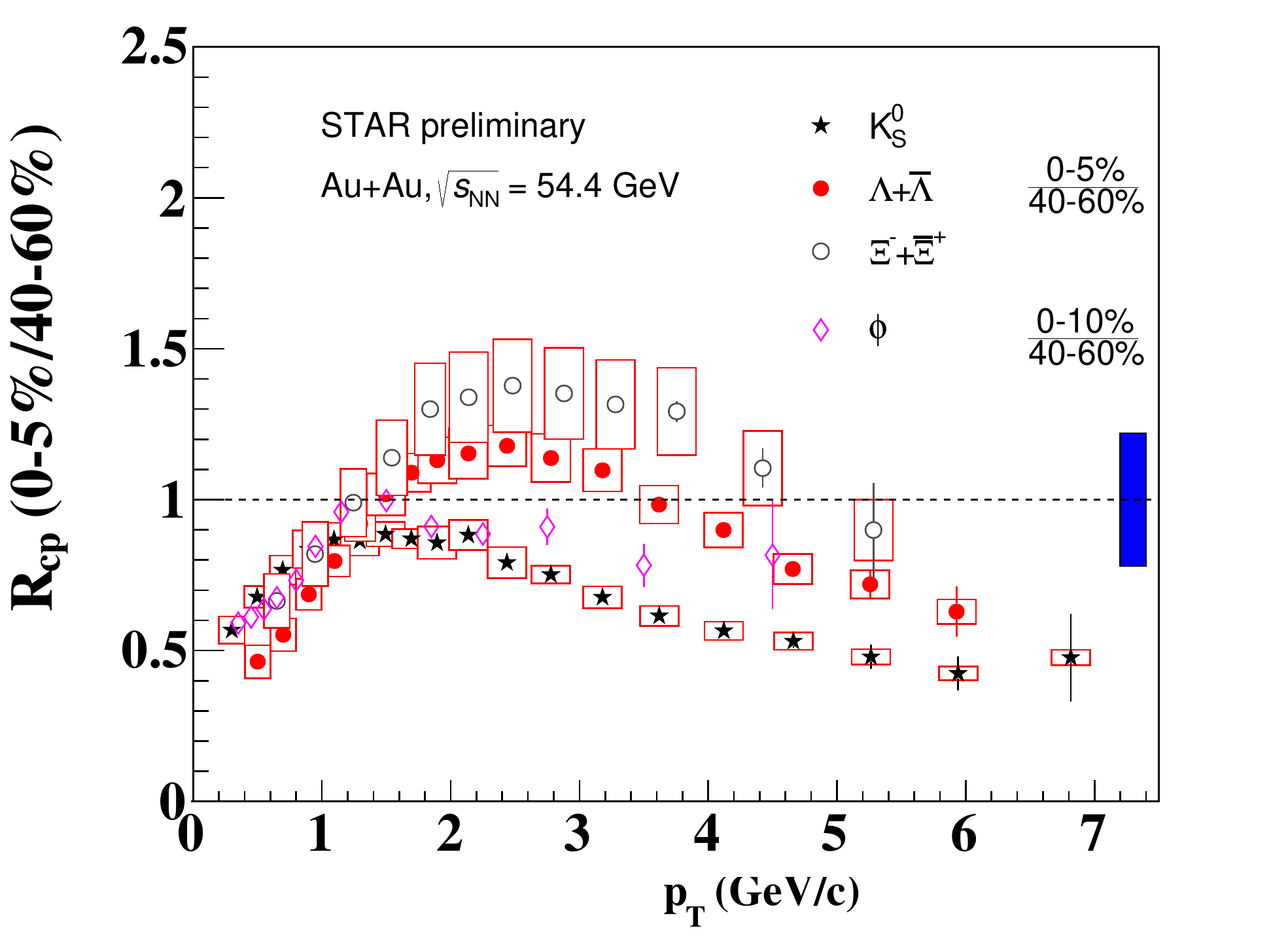}
\caption{The nuclear modification factor, $R_{CP}$ is plotted as a function of $p_{T}$ at $\sqrt{s_{NN}}$ = 54.4 GeV for $K_{S}^{0}$, $\Lambda + \bar{\Lambda}$, $\Xi^{-} + \bar{\Xi^{+}}$ and for $\phi$. The vertical lines represent the statistical error bars and the rectangular boxes represent the systematic error bars. Systematic errors are not shown for $\phi$ mesons. The blue box represents the uncertainty from the determination of the number of binary collisions.}
\label{fig-3}       
\end{figure}

Figure~\ref{fig-3} shows the nuclear modification factor ($R_{CP}$) at $\sqrt{s_{NN}}$ = 54.4 GeV, which is calculated by taking ratio of particle yields in 0-5\% and 40-60\% for $K_{S}^{0}$, $\Lambda + \bar{\Lambda}$, $\Xi^{-} + \bar{\Xi^{+}}$ and for $\phi$ mesons. The ratio shows a strong suppression at high $p_{T}$ indicating partonic energy loss inside the medium.


\begin{figure}[h]
\centering
\includegraphics[width=6cm,clip]{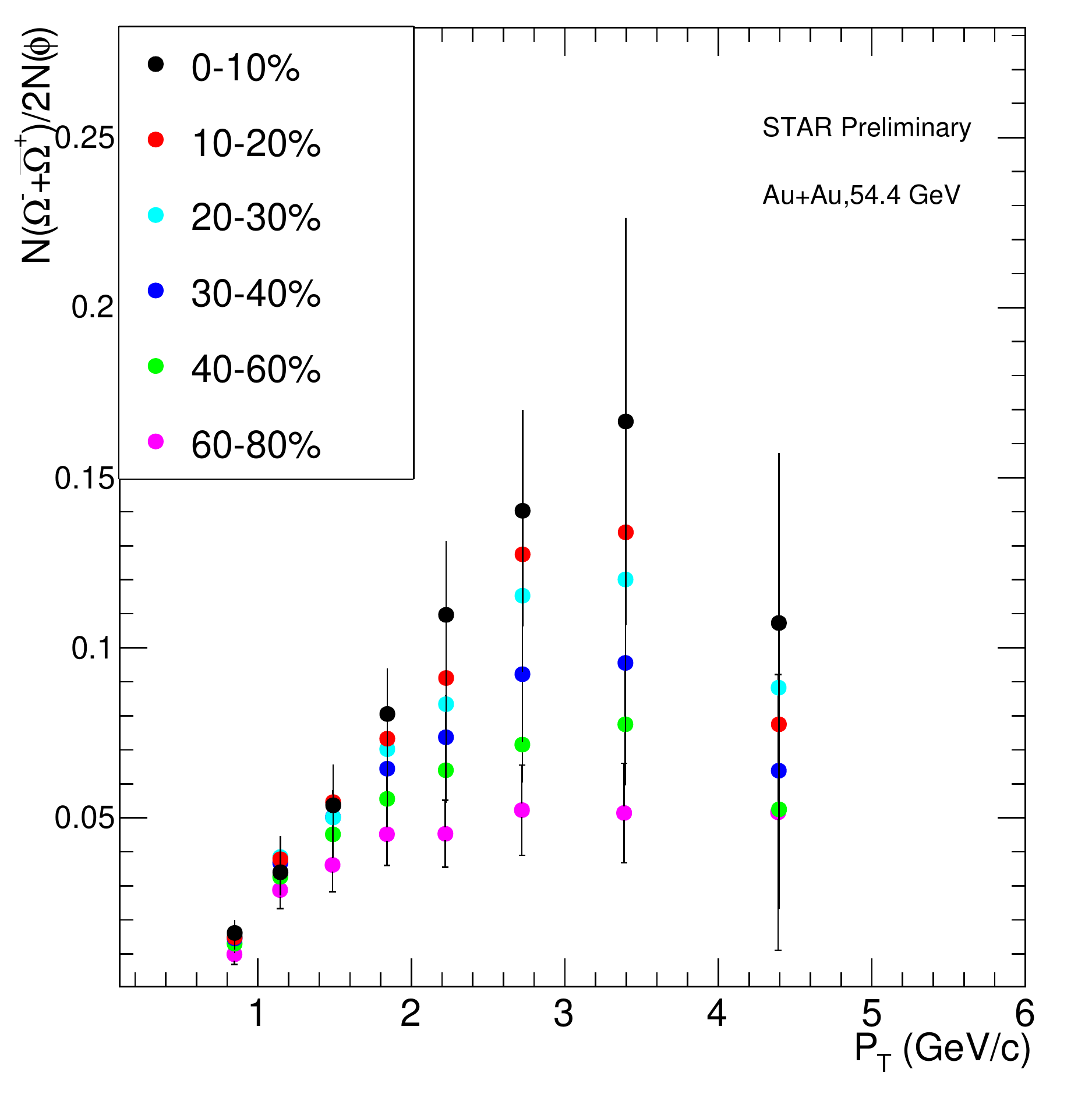}
\caption{$N[\Omega + \bar{\Omega}^{+}]/2N[\phi]$ is plotted as a function of $p_{T}$ for different centrality classes at  $\sqrt{s_{NN}}$ = 54.4 GeV. Only statistical error bars are shown.}
\label{fig-4}       
\end{figure}

\begin{figure}[h]
\centering
\includegraphics[width=7cm,clip]{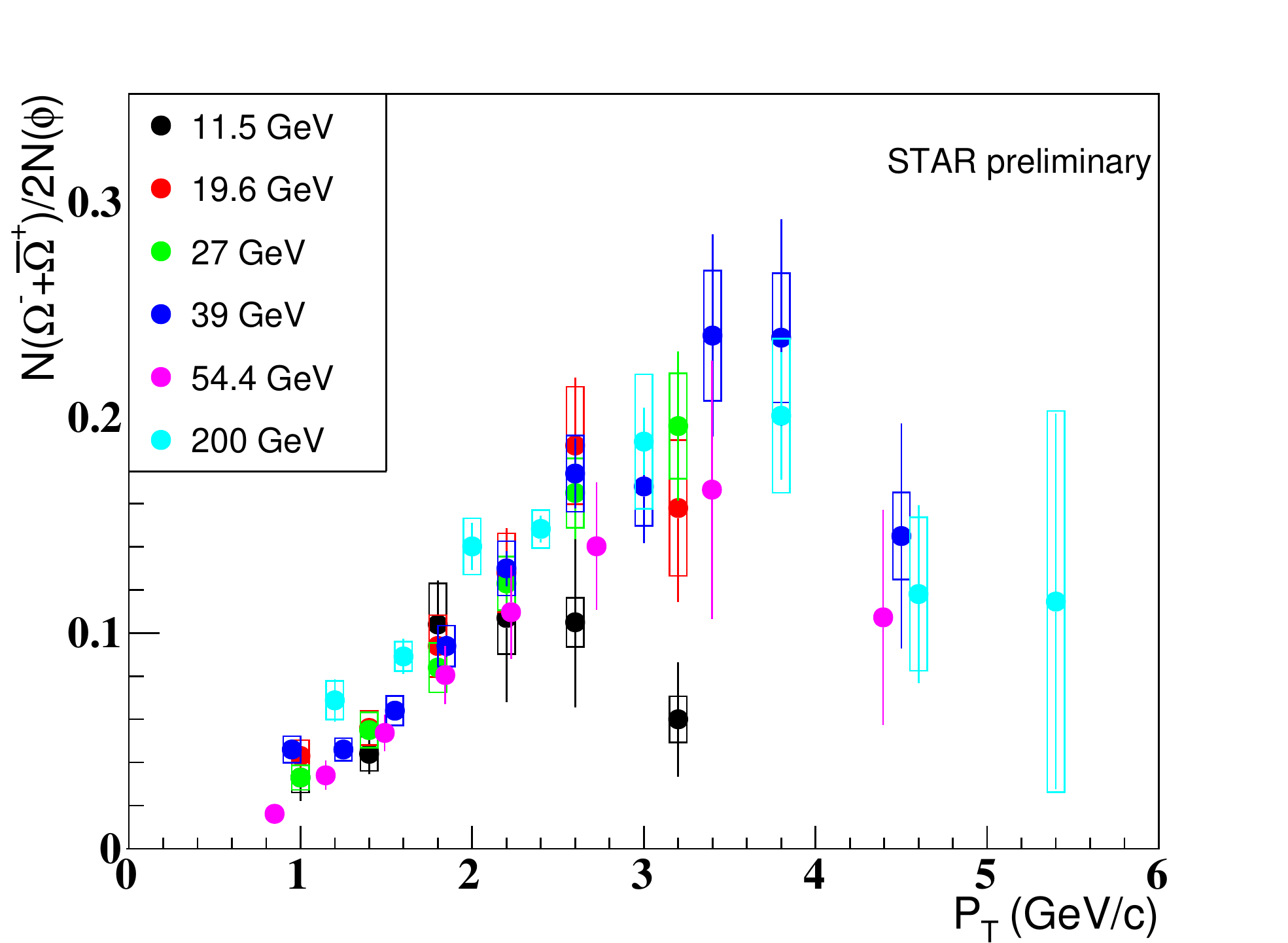}
\caption{$N[\Omega + \bar{\Omega}^{+}]/2N[\phi]$ ratio vs. $p_{T}$ for 0-10\% centrality at  $\sqrt{s_{NN}}$ = 54.4 GeV is compared with other STAR energies for most central collisions. For 54.4 GeV only statistical errors are shown.}
\label{fig-5}       
\end{figure}

The ratio, $N[\Omega + \bar{\Omega}^{+}]/2N[\phi]$ is calculated for different centralities as shown in Fig.~\ref{fig-4}. The ratio shows an enhancement at intermediate $p_{T}$ and the enhancement is larger for central collisions than peripheral collisions which suggests the idea of quark coalescence hadronization~\cite{Ref4}. The ratio at 54.4 GeV is compared with other STAR energies~\cite{Ref5} for most central collisions as shown in Fig.~\ref{fig-5}. $N[\Omega + \bar{\Omega}^{+}]/2N[\phi]$ ratio shows the same trend and is consistent with other energies except for 11.5 GeV data where the larger statistical error bar restrict us from making any conclusion.

\begin{figure}[h]
\centering
\includegraphics[width=10cm,clip]{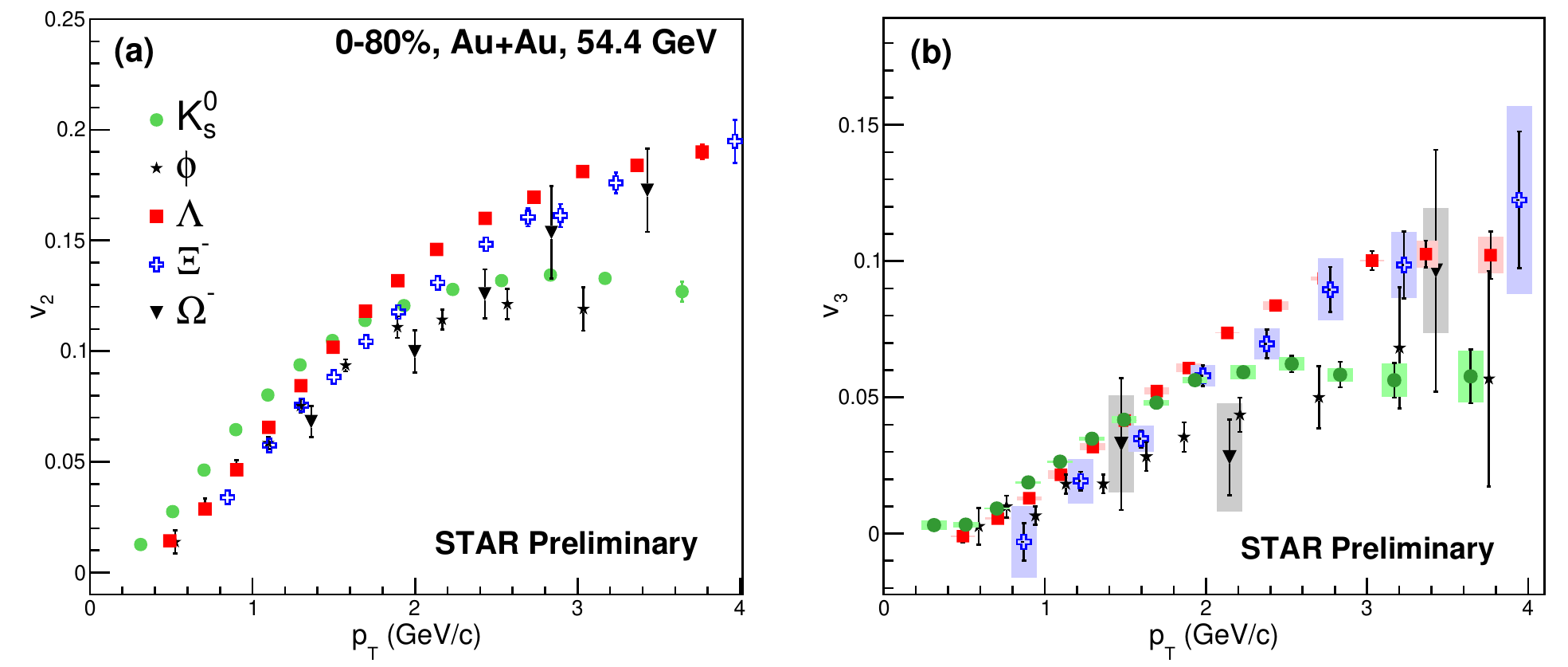}
\caption{Left panel shows the $v_{2}$ as a function of $p_{T}$ for 0-80\% centrality at $\sqrt{s_{NN}}$ = 54.4 GeV for $K^{0}_{S}$, $\Lambda$, $\phi$, $\Xi^{-}$ and $\Omega^{-}$. The right panel shows the same for $v_{3}$. The vertical bars here are the statistical error bars and the shaded boxes are the systematic error bars.}
\label{fig-6}       
\end{figure}

\begin{figure}[h]
\centering
\includegraphics[width=10cm,clip]{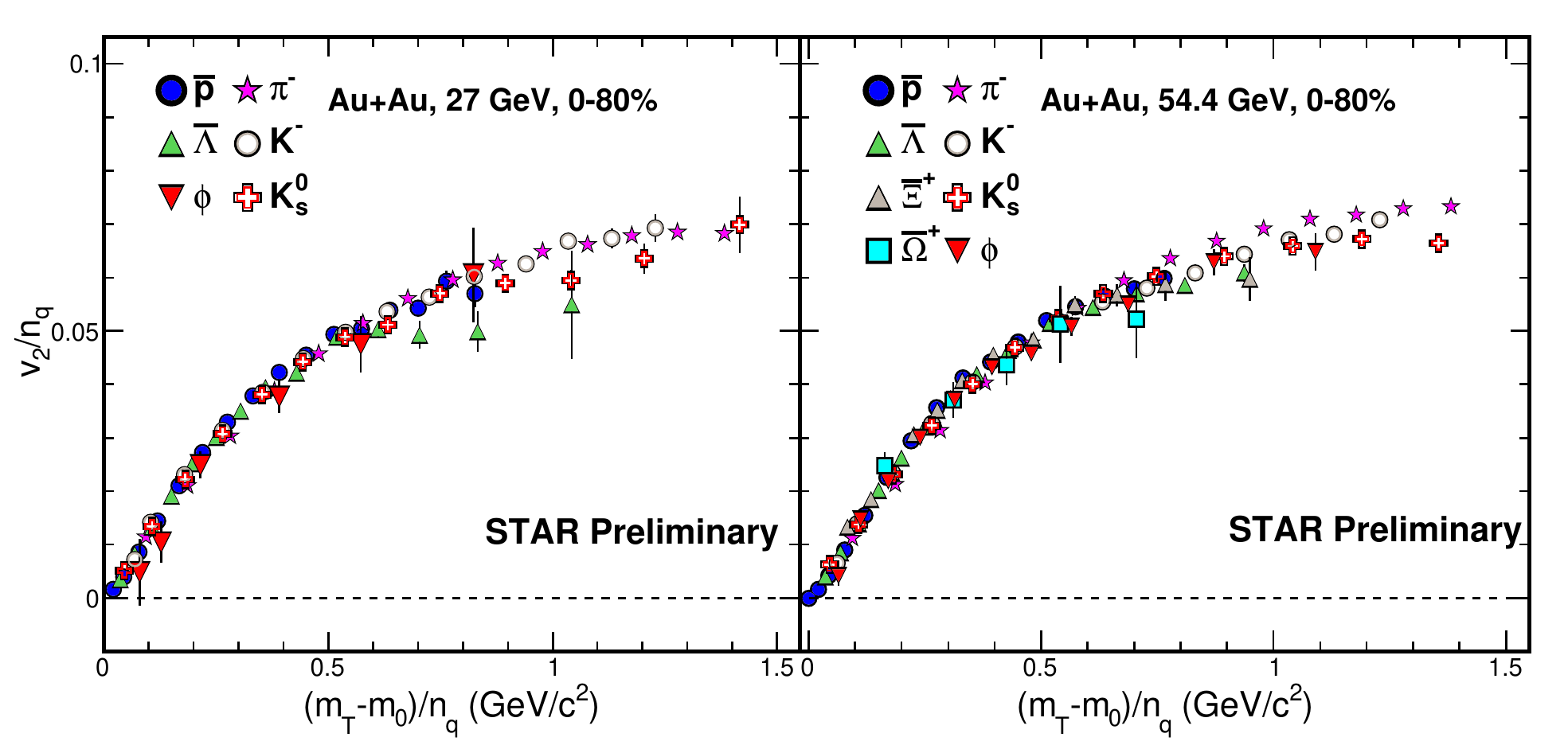}
\caption{The left panels shows the NCQ scaled $v_{2}$ as a function of transverse kinetic energy at $\sqrt{s_{NN}}$ = 27 GeV for 0-80\% centrality. The right panel shows the same for $\sqrt{s_{NN}}$ = 54.4 GeV. Only statistical error bars are shown. }
\label{fig-7}       
\end{figure}

\begin{figure}[h]
\centering
\includegraphics[width=7cm,clip]{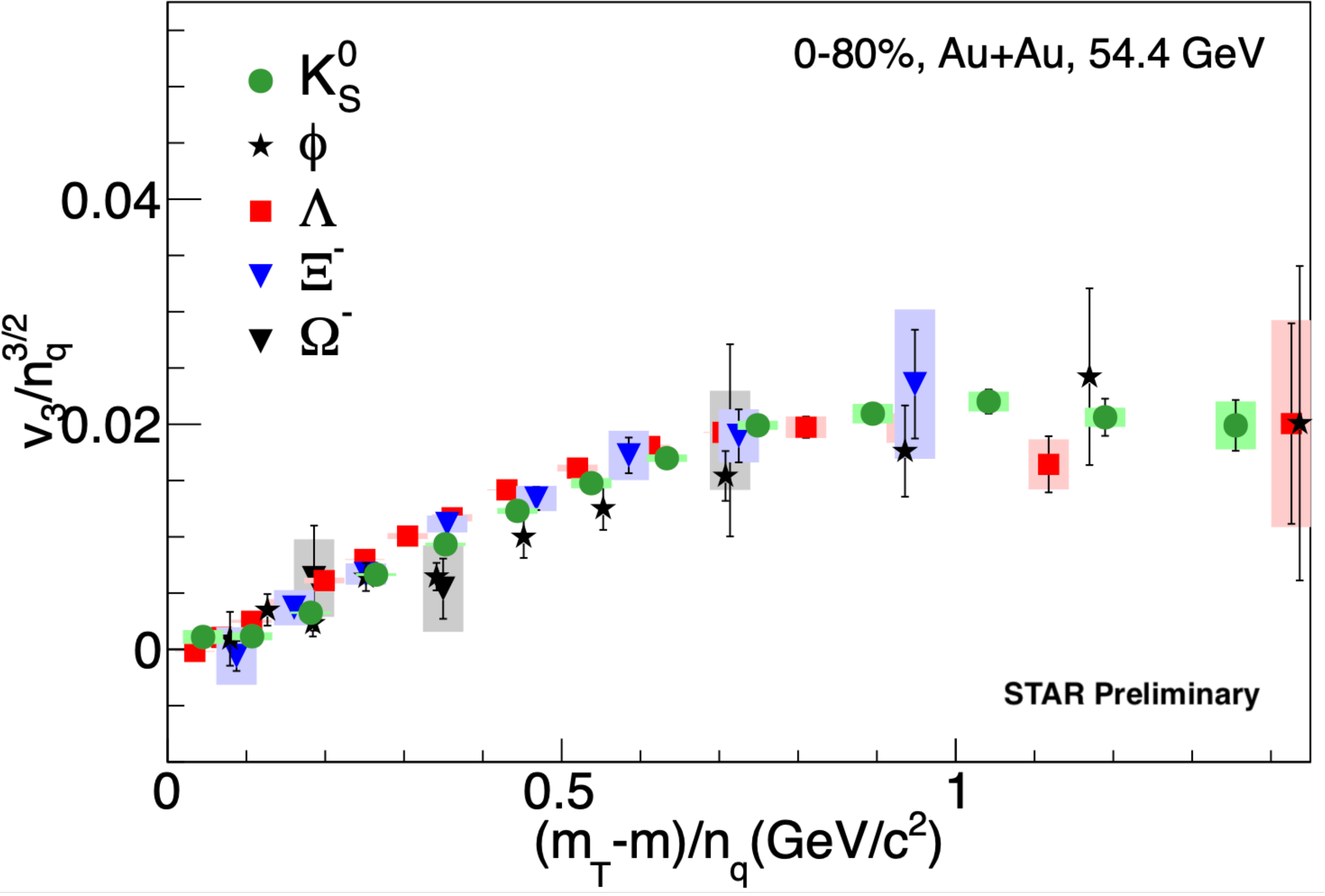}
\caption{NCQ scaled $v_{3}$ is plotted as a function of transverse kinetic energy at $\sqrt{s_{NN}}$ = 54.4 GeV for 0-80\% centrality. The vertical bars here are the statistical error bars and the shaded boxes are the systematic error bars.}
\label{fig-8}       
\end{figure}

The elliptic ($v_{2}$) and triangular ($v_{3}$) flow coefficients are measured for the strange and multi-strange hadrons for minimum bias events at $\sqrt{s_{NN}}$ = 54.4 GeV as shown in Fig.~\ref{fig-6}. A mass ordering is observed at low $p_{T}$ ($<$ 2 GeV/$c$), where the particles with smaller mass show a higher flow which is due to the radial flow of the system during the QGP phase. A particle type dependence is observed in both $v_{2}$ and $v_{3}$ at high $p_{T}$ region. We have studied the NCQ scaled $v_{2}$ as a function of NCQ scaled transverse kinetic energy at $\sqrt{s_{NN}}$ = 27 and 54.4 GeV for 0-80\% centrality as shown in Fig.~\ref{fig-7}. The NCQ scaled $v_{2}$ falls on a single curve for all the hadron, which indicates that the collectivity is developed during the initial QGP phase of the system. Similar behaviour is observed in case of NCQ scaled $v_{3}$ at $\sqrt{s_{NN}}$ = 54.4 GeV as shown in Fig.~\ref{fig-8}.

\section{Summary}
In summary, we have measured the $p_{T}$ spectra of strange hadrons, $K^{0}_{S}$, $\Lambda$, $\bar{\Lambda}$, $\phi$, $\Xi^{-}$, $\bar{\Xi}^{+}$, $\Omega^{-}$, $\bar{\Omega}^{+}$ at midrapidity in Au+Au collisions at $\sqrt{s_{NN}}$ = 54.4 GeV. The nuclear modification factor, $R_{CP}$ for the above mentioned particles has been measured which shows a strong suppression at high $p_{T}$. The baryon to meson ratio, $N[\Omega + \bar{\Omega}^{+}]/N[\phi]$ shows an enhancement at intermediate $p_{T}$ which is higher for central collisions than that for peripheral collisions. The anisotropic flow coefficients, $v_{2}$ and $v_{3}$, have been measured as a function of $p_{T}$  in 0-80\% centrality for all the strange particles mentioned above. A mass ordering is observed at low $p_{T}$ and a baryon to meson separation is observed at high $p_{T}$. The NCQ scaling founds to hold for both $v_{2}$ and $v_{3}$.

\end{document}